\documentclass[aps,prl,twocolumn,showpacs]{revtex4}
\usepackage{ifthen}
\newcommand{\usepdfs}{false}
\ifthenelse{\equal{\usepdfs}{true}}{
  \usepackage[pdftex]{graphicx}
}
{
  \usepackage[dvips]{graphicx}
}
\ifthenelse{\equal{\usepdfs}{true}}{
  \newcommand{\ifig}[2][]{\includegraphics[#1]{#2.pdf}}
}
{
  \newcommand{\ifig}[2][]{\includegraphics[#1]{#2.eps}}
}

\usepackage{amssymb,amsmath,amsfonts,hyperref}
\usepackage{wasysym}
\usepackage{latexsym}

\newcommand{\ket}[1]{| #1 \rangle}
\newcommand{\bra}[1]{\langle #1 |}

\newcommand{\be}{\begin{equation}}
\newcommand{\ee}{\end{equation}}

\newcommand{\Jx}{J_{\perp}}

\newcommand{\cdaf}{CdCr$_2$O$_4$}
\newcommand{\hgaf}{HgCr$_2$O$_4$}

\begin{document}

\title{Quantum effects in a half-polarized pyrochlore antiferromagnet} \date{\today}

\author{Doron L. Bergman$^1$, Ryuichi Shindou$^1$, Gregory
  A. Fiete$^{1,2}$, and Leon Balents$^1$}
\affiliation{${}^1$Department of Physics, University of California,
  Santa Barbara, CA 
93106-9530\\$^2$Kavli Institute for Theoretical Physics, University of 
California, Santa Barbara, CA 93106-4030}

\begin{abstract}
  We study quantum effects in a spin-$3/2$ antiferromagnet on the
  pyrochlore lattice in an external magnetic field, focusing on the
  vicinity of a plateau in the magnetization at half the saturation
  value, observed in \cdaf\, and \hgaf.  Our theory, based on quantum
  fluctuations, predicts the existence of a symmetry-broken state on the
  plateau, even with only nearest-neighbor microscopic exchange.  This
  symmetry broken state consists of a particular arrangement of spins
  polarized parallel and antiparallel to the field in a 3:1 ratio on
  each tetrahedron.  It quadruples the lattice unit cell, and reduces
  the space group from $Fd\overline{3}m$ to $P4_332$.  We also predict
  that for fields just above the plateau, the low temperature phase has
  transverse spin order, describable as a Bose-Einstein condensate of
  magnons.  Other comparisons to and suggestions for experiments are
  discussed.
\end{abstract}
\date{\today}
\pacs{75.10.-b,75.10.Jm,75.25.+z}

%75.10.Jm Quantized spin models
%75.10.-b General theory and models of magnetic ordering (see also 05.50.+q Lattice theory and statistics)
%75.25.+z Spin arrangements in magnetically ordered materials (including neutron and spin-polarized electron studies, synchrotron-source x-ray scattering, etc.) (for devices exploiting spin polarized transport, see 85.75.-d)
%75.60.-d Domain effects, magnetization curves, and hysteresis

%\email{doronber@physics.ucsb.edu}

\maketitle

%\tableofcontents

Frustrated quantum magnets, in which many symmetry-unrelated states are
classically degenerate, are a fascinating venue in which to observe
emergent phenomena. In most frustrated materials,
the degeneracy is lifted classically, by lattice distortions
(Jahn-Teller, spin-Peierls\cite{TMS}), or longer-range (e.g. dipolar) interactions\cite{Melko}.  
More intriguing is the possibility that the degeneracy can be removed by
quantum fluctuations.  In such a ``quantum order by disorder'' scenario,
the system would pick a ground state by maximally delocalizing among
many degenerate classical states, thereby minimizing its quantum zero
point energy.  In toy models, this can lead to exotic ordered states
(such as valence bond solids), or even ``quantum spin liquid''\cite{Hermele} states,
where the classical degeneracy is split {\sl without any symmetry
  breaking}.  Experimentally, clean signatures of the lifting of degeneracy by
quantum fluctuations have, however, remained elusive, presumably because
of the dominance of the classical mechanisms discussed above.

\begin{figure}[htb]
        \centering
\ifig[width=3.0in]{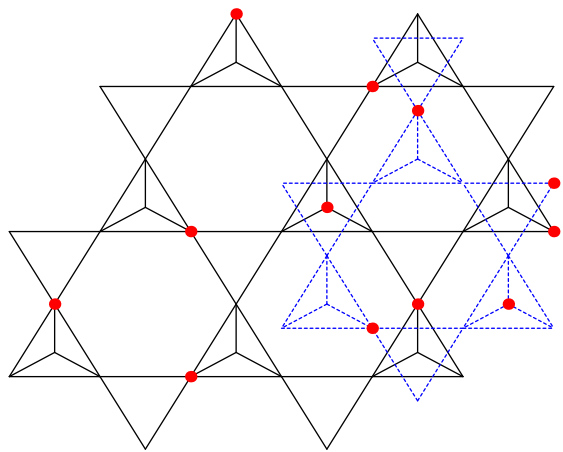}
\caption{(Color online) Planar projection of the ``R'' state, stabilized by quantum
  fluctuations. Dots indicate minority spins.  The dashed lines are
  parts of the next layer of tetrahedra}
\label{fig:charge1}
\end{figure}

In this paper we describe a theoretical study of quantum effects in
the insulating chromium oxides, \cdaf\, and \hgaf,
in which magnetic spin-$3/2$ Cr$^{3+}$ ions form a pyrochlore lattice
(a network of corner-sharing tetrahedra).  Because of the half-filled
$t_{2g}$ magnetic levels, these materials lack orbital degeneracy,
and are also very isotropic magnetically.  An appropriate minimal
theoretical model is thus the 
nearest-neighbor pyrochlore lattice Heisenberg antiferromagnet, with
Hamiltonian
\begin{equation}
  \label{H_b}
  {\mathcal H} = 
  J \sum_{\langle i j \rangle} {\bf S}_i \cdot {\bf S}_j
  - H  \sum_j S^z_j
  \; ,
\end{equation}
where ${\bf S}_i$ is a spin-$3/2$ operator on site $i$ of the pyrochlore
lattice, $\langle ij\rangle$ indicates the sum is over nearest-neighbor
bonds, and we have taken the field $H$ to point along the $z$-axis.
Most intriguingly, these materials display a robust low-temperature
magnetization plateau in an applied magnetic field, with magnetization
well-quantized at half its saturation value\cite{Ueda:prl05}.  While
there has been a great deal of prior theoretical work on quantum effects
in zero magnetic
field\cite{Hizi:05,tchern:prb03}, the
connection to experiment seems unclear.  Here we will instead focus on
the physics on and near the plateau.  First, we predict a magnetic
structure on the plateau shown in Fig.~\ref{fig:charge1}, with space
group $P4_332$, which we call the ``R'' state.  Second, we predict the
existence of XY order (i.e.  transverse to the field) at low temperature
for fields just above the plateau.  If quantum fluctuations dominate
over second neighbor exchange, the transverse order is {\em
  ferromagnetic}.

These conclusions obtain both for Eq.~(\ref{H_b}) {\sl and} a modified
model including spin-lattice coupling to local Einstein bond phonons,
which Penc {\sl et al.}\cite{Penc:prl04} have suggested play an
important role.  While we agree such coupling likely supports the large
{\sl width} of the plateau, our conclusions with respect to the {\sl
  ordering symmetry} are essentially independent of it.  For clarity of
presentation, we therefore suppress spin-lattice coupling in the
following.

Our approach relies on the substantial polarization in the vicinity of
the plateau, which indicates that, on average, the transverse components
of the spins are reduced.  Moreover, the existence of a plateau with
zero differential susceptibility $\chi= \partial m^z/\partial h = 0$
at $T=0$ implies the absence of transverse ordering, $\langle
S_i^\pm\rangle=0$.  We therefore split the Hamiltonian into terms
involving longitudinal and transverse spin components, and {\sl treat
  the latter perturbatively} (though we will work to rather high
orders).  Specifically, ${\mathcal H}'={\mathcal H}_0 + {\mathcal
  H}_1$, with
\begin{eqnarray}
\label{H_0} &
{\mathcal H}_0 & = \frac{J_z}{2} \sum_t \left[ (S_t^z - h)^2  - h^2
\right] - J_z \sum_i \left( S_i^z \right)^2 
\; ,
\\ \label{H_1} &
{\mathcal H}_1 & = \frac{\Jx}{2} \sum_{\langle i j \rangle} \left(
  S_i^+ S_j^- + h.c. \right)\;, 
\end{eqnarray}
where $S_i^{\pm}=S_i^x \pm i S_i^y$ are the usual ladder operators,
$S_t^z = \sum_{j \in t} S_j^z$ is the sum of longitudinal spin
components on a tetrahedron labeled by $t$, and we have introduced the
dimensionless magnetic field $h = H/2J$ (we take $h>0$ without loss of
generality).  For bookkeeping purposes, we have written distinct
longitudinal and tranverse exchange couplings $J_z,\Jx$, though we are
ultimately interested in $J_z=\Jx=J$.  We note that ``easy axis''
perturbation theory in $\alpha\equiv \Jx/J_z$ becomes increasingly
accurate at larger $h$, and that the symmetry of ${\mathcal H}'$ is
preserved for any $\alpha>0$.  This
gives us further confidence in the {\sl qualitative} predictions of
easy-axis perturbation theory.  

We now sketch our derivation of an {\sl effective model} describing
the quantum mechanical lifting of the classical degeneracy.  Consider
first the 0$^{\textrm{th}}$ order Hamiltonian, Eq.~(\ref{H_0}).  It is
straightforward to see that ${\mathcal H}_0$ has three types of ground
states, depending upon $h$: for $h<1.5$, each tetrahedron has two
spins with $S_i^z=+\frac{3}{2}$ and two spins with
$S_i^z=-\frac{3}{2}$; for $1.5 < h < 4.5$, the half-magnetized region,
each tetrahedron has 3 ``majority'' spins with $S_i^z=+\frac{3}{2}$
and one ``minority'' spin $S_i^z=-\frac{3}{2}$; and all spins are
fully polarized for $h>4.5$.  Therefore, this simplistic view predicts
3 magnetization plateaus exist -- zero, half and full magnetization.

Focusing first on the plateau of interest, the
0$^{\textrm{th}}$ order ``3:1'' ground states are extensively
degenerate, due to the freedom to locate the minority spin of each
tetrahedron.  We may view the transverse terms in Eq.~(\ref{H_1}) as
inducing ``quantum fluctuations'' into these states; in particular,
the action of ${\mathcal H}_1$ transfers a total spin $\Delta S^z =
\pm 1$ between nearest neighbor sites.  Technically, this must be
analyzed by Degenerate Perturbation Theory (DPT).

We define the
projection operator, ${\mathcal P}$, onto the ground state plateau
subspace.  Consider any exact eigenstate $|\Psi\rangle$.  Its
projection $|\Psi_0\rangle={\mathcal P}|\Psi\rangle$ satisfies the
``effective Schr\"odinger equation''
\begin{equation}
  \label{eq:effshrod}
   \left[ E_0 + 
{\mathcal P} {\mathcal H}_1 \sum_{n=0}^{\infty} {\mathcal G}^n 
{\mathcal P} \right] |\Psi_0\rangle = E |\Psi_0\rangle,
\end{equation}
where the resolvent operator ${\mathcal G} = \frac{1}{E - {\mathcal H}_0} \left( 1 -
  {\mathcal P} \right) {\mathcal H}_1 $ is linear in $\Jx$.  Because the
resolvent contains the exact energy $E$, Eq.~(\ref{eq:effshrod}) is
actually a non-linear eigenvalue problem.  However, to any given order
of DPT, $E$ may be expanded in a series in $\Jx$ to obtain an equation
with a true Hamiltonian form within the degenerate manifold.  Once
$|\Psi_0\rangle$ and $E$ are known, the full wavefunction can be reconstructed
as $\ket{\Psi} = (1-{\mathcal G})^{-1} \ket{\Psi_0} =
\sum_{n=0}^{\infty} {\mathcal G}^n \ket{\Psi_0}$. 

To understand the ultimate nature of the ordered state on the plateau,
we must carry out DPT at least to the lowest order at which the
degeneracy is lifted.  Our approach will be to derive the lowest order
{\sl diagonal} (in the $S_i^z$ basis) and {\sl off-diagonal} terms in
$\Jx$ which remove the degeneracy.  First, note that any {\sl
  off-diagonal} term must flip spins in such a way as to preserve the
3:1 constraint on each tetrahedron.  This can only be accomplished by
flipping spins around a non-trivial closed loop on the pyrochlore
lattice.  The smallest such loop involves flipping spins on 3 different
bonds, and flipping a spin from $\pm \frac{3}{2}$ to $\mp \frac{3}{2}$
requires ${\mathcal H}_1$ to act 3 times, so off-diagonal processes
occur first at 9$^{\mathrm th}$ order (for general $S$, they occur at
$O(\Jx^{6S})$)!  Diagonal processes can occur sooner.  The first-order
splitting vanishes, since the action of ${\mathcal H}_1$ creates two
tetrahedra with spin $S_t^z \neq 3$.  The energy of the plateau state 
does shift relative to other levels at second order.  Although it is 
more non-trivial to see, in fact, splitting {\sl within} the classical manifold
does not occur until {\sl sixth} order (this is true for any $S$)!  The
reason for this is somewhat involved\cite{longpaper}, but follows from the {\sl
  locality} of ${\mathcal H}_1$ and the strong 3:1 constraint.  

By straightforward but somewhat involved calculations, one can compute
the lowest-order diagonal and off-diagonal contributions to the
effective Hamiltonian, ${\mathcal H}_{\rm eff}$, explicitly.  We find
\begin{equation}
\label{H_eff}
{\mathcal H}_{\textrm{eff}} = \alpha^6 J_z \sum_P \hat{\mathcal E}_P
- c \alpha^9 J_z \sum_P \left( \ket{\hexagon_A} \bra{\hexagon_B} + {\rm h.c.} \right) 
\; ,
\end{equation}
where the sums are over all hexagonal plaquettes
$P$, and the states $\ket{\hexagon_A},\ket{\hexagon_B}$ represent the
two ``flippable'' configurations (A/B) with alternating majority and
minority spins around the plaquette. The operator $\hat{\mathcal E}_P$
is a diagonal operator representing the 6$^{\mathrm{th}}$ order energy
splitting of plaquette $P$.  It takes the values $\hat{\mathcal E}_P
=(0,-\frac{144}{25},-\frac{27}{100},-\frac{801}{500},-\frac{9801}{6250})\approx(0,-5.76,-0.27,-1.60,-1.57)$,
respectively for the configurations
$(\uparrow\uparrow\uparrow\uparrow\uparrow\uparrow,\downarrow\uparrow\downarrow\uparrow\downarrow\uparrow,\downarrow\uparrow\uparrow\uparrow\uparrow\uparrow,\downarrow\uparrow\downarrow\uparrow\uparrow\uparrow,\downarrow\uparrow\uparrow\downarrow\uparrow\uparrow)$
(and cyclic permutations), where $\uparrow/\downarrow$ denote
majority/minority spin states, respectively.  This diagonal interaction
is equivalent to a combination of further neighbor Ising
exchange couplings, plus an additional 3-spin interaction term; the
plaquette representation is, however, more convenient and simpler.
The constant $c=\frac{53178588}{12371645}\approx 4.3$.

The Hamiltonian in Eq.~(\ref{H_eff}) (acting within the 3:1 Hilbert
space) is the basis for our discussion of the ordered state on the
plateau.  Inspection of $\hat{\mathcal E}_P$ shows that the A/B
configurations are significantly more favorable than all others.  A
natural approximation is then to ignore the energy differences between
other configurations, which gives
\begin{eqnarray}
  \label{eq:Happrox}
  {\mathcal H}_{\textrm{QDM}} & = & V \sum_P \left(
  \ket{\hexagon_A} \bra{\hexagon_A} + \ket{\hexagon_B} \bra{\hexagon_B}
\right) \nonumber \\ &&  - K \sum_P \left( \ket{\hexagon_A} \bra{\hexagon_B} + h.c.
\right), 
\end{eqnarray}
where, extrapolating to the physical limit $\alpha\rightarrow 1$, we
have $V\approx -5.76 J$, $K=c J \approx 4.3 J$, while in the strict small
$\alpha$ limit, $V/K \rightarrow -\infty$.  Eq.~(\ref{eq:Happrox}) is
useful because it maps directly to a type of ``Quantum Dimer Model''
(QDM)\cite{RK:prl88}.  In particular, the pyrochlore sites can be mapped onto
the links of a diamond lattice, whose sites lie in the centers of
pyrochlore tetrahedra.  A minority spin can be represented as a ``dimer''
occupying the corresponding link, and the 3:1 states become
non-overlapping dimer coverings of the diamond lattice.
%Furthermore, because the diamond
%lattice is bi-partite, the links may be oriented (from e.g. ``up'' to
%``down'' pyrochlore tetrahedra centers), and the dimers thereby
%associated with a vector ``electric field'' variable.  In this way,
%the ground state Hilbert space is mapped to that of a quantum dimer
%model and equivalent compact U(1) gauge theory.  
The $V$ and $K$ terms map directly to standard ``potential'' and
``kinetic'' terms for these dimer configurations\cite{RK:prl88}.

QDMs of this form have been considered on a number of two and
three-dimensional
lattices\cite{RK:prl88,Huse}.
Fixing $K>0$, the ground state depends upon the dimensionless parameter
$v=V/K$, and a few general conclusions can be drawn.  The
Rokhsar-Kivelson point, $v=1$, is exactly soluble, and demarks a
boundary between ordering into configurations with no A/B plaquettes
(for $v>1$) and, for three-dimensional bipartite lattices like the
diamond, a {\sl spin liquid} phase with no broken symmetry (for
$v_{c1}<v \lesssim 1$\cite{Hermele}, with some critical coupling $v_{c1}$ which is
usually positive).  For $v < v_{c2} < v_{c1}$, the ground state is again
ordered, and adiabatically connected to the diagonal ground state
selected by the diagonal $V<0$ term alone.  For the examples (square,
honeycomb, and triangular lattices) in which the phase diagram has been
studied by quantum Monte Carlo
methods\cite{QDM}, the critical
coupling $v_{c2}$ is either slightly negative or even positive,
$v_{c2}>-0.2$.  The persistence of the ``diagonal'' state to small $v$
can be readily understood.  For $v\rightarrow-\infty$, the energy is
minimized by the diagonal state which contains the maximum density of
flippable A/B hexagons.  Such a configuration is {\sl also} the one most
connected by the $K$ term to other 3:1 states.  Hence, by ``order by
disorder'' reasoning, an appropriate superposition ``centered'' (in
Hilbert space) around the diagonal ground state above is energetically
favored by {\sl both} terms.  Explicit variational wavefunctions for
${\mathcal H}_{\rm eff}$ in this spirit will be described in a future
long publication.  What or how many intervening state(s?) might occur
for $v_{c2}<v<v_{c1}$ is not known, and it is even possible that there
might be no intermediate state, and instead a direct transition from the
diagonal ordered state to a spin liquid at $v=v_{c2}=v_{c1}$.

$ $From this reasoning, in the extrapolation to $\alpha=1$, which
gives $v\approx -1.3$, we expect the ground state selected by quantum
fluctuations to have the same symmetry as that of the easy-axis limit
$v\rightarrow-\infty$.  In particular, for the QDM in this limit, we
require the 3:1 configuration with the highest density of A/B
hexagons.  Because these hexagons overlap, and the 3:1 configuration
space is highly constrained, this is not a trivial problem.  It is
instructive to examine a single primitive unit cell of the pyrochlore
lattice, which is a region of space enclosed by 4 hexagonal
plaquettes.  By careful inspection, and making use of the 3:1 constraint,
one finds that at most one of these hexagons can be of A/B type.
Therefore the maximum fraction of A/B hexagons is $\frac{1}{4}$.

By extensive analysis\cite{longpaper}, we have found one single
candidate, shown in Fig.~\ref{fig:charge1}, which saturates this upper
bound on the density of A/B hexagons, and moreover, gives the lowest
energy that we have been able to find for the {\sl full} diagonal term
($\sum_P \hat{\mathcal E}_P$) with {\sl all} plaquette energies
included.  Having the maximal density of potential Resonating
plaquettes, we call it the ``R'' state.  It has a magnetic unit cell
consisting of $4$ structural pyrochlore unit cells, and has the
symmetries of the $P4_332$ space group -- particularly significant is
that this state preserves all rotation symmetries of the pyrochlore
lattice (up to translations).  This implies that, contrary to the
suggestion in Ref.~\cite{Ueda:prl05}, only {\sl isotropic}
magnetostriction is expected for the R state, and not a rhombohedral
distortion along a $\langle 111 \rangle$ axis.  We note that a Landau
theory analysis predicts the symmetry-breaking transition from the
$P4_332$ space group (R state) to $Fd\overline{3}m$ (pyrochlore
symmetry) is first order\cite{longpaper}, in agreement with the
experimental conclusions\cite{Ueda:new}.

We now turn to other effects of quantum fluctuations.  Quantum effects lead
to a {\sl spin gap}, $\Delta$, on the plateau, and consequently an activated
temperature dependence, $M^z(T)-M^z(0)
\propto \exp(-\Delta(h)/k_{\scriptscriptstyle B}T)$.  Another quantum
effect is a suppression of the local ``staggered'' moment, even at
$T=0$.  To leading 
order in DPT, the difference of the moment on majority
and minority sites is suppressed to $\langle S_{\rm max}^z\rangle
-\langle S_{\rm min}^z\rangle < 2.2$, below the classical value
$\langle S_{\rm max}^z\rangle -\langle S_{\rm min}^z\rangle = 3$.

Experiments have studied the low-temperature behavior on {\sl exiting}
the plateau at lower and higher fields, $h_{c1},h_{c2}$, respectively.
The lower edge of the plateau for both \cdaf$ $\cite{Ueda:prl05} 
and \hgaf$ $\cite{Ueda:new} shows a jump 
in magnetization, which then decreases linearly down to zero
magnetic field, suggesting the low-field state is adiabatically
connected with the zero field state.  The higher field edge, observed
only in \hgaf, shows a continuous transition out of the plateau, and a
continuous evolution with no further jumps up to the highest fields
measured.  There is however some indication of a ``kink'', possibly
related to the expected lattice symmetry restoration, for fields above the
plateau but below full saturation.

The first order/continuous nature of the transitions at the lower/upper
edges can be explained classically\cite{Penc:prl04}, and occurs already
in the trivial problem of a {\sl single} classical tetrahedron.
Classically, however, above the plateau, though the spins become
non-collinear, the {\sl ordering} of the transverse moments, if present
at all, is critically dependent upon the nature of further neighbor
interactions.

By contrast, it is determined in the quantum theory even for the
nearest-neighbor model.  In general, a continuous quantum transition off
the plateau should be described by the ``Bose-Einstein Condensation''
(BEC) of a single magnon excitation with $\Delta S^z = +1$, above the R
state, whose gap vanishes at the plateau edge.  The resulting state
above the plateau has therefore transverse spin order, $\langle
S_i^\pm\rangle\neq 0$, in agreement with classical expectations.  The
precise spatial arrangement of the transverse ordering is determined by
the symmetry of the lowest energy magnon, which must form an irreducible
representation (irrep) of $P4_332$.  In the easy axis DPT, the magnon is
formed from a superposition of states with a single minority spin changed
from $S_i^z=-\frac{3}{2}$ to $S_i^z=-\frac{1}{2}$.  At $O(\alpha^2)$,
the magnon can hop between minority sites, and crucially, the effective
hopping amplitude is {\sl negative}, despite $J>0$, because it occurs by
virtual hopping through a majority site.  The minimum energy magnon
state then is a {\sl uniform} plane wave (trivial irrep), which implies
the transverse spin order is {\sl ferromagnetic}.  That is, for
$h\gtrsim h_{c2}$ the system has {\sl the same space group symmetry as
  the R state}.  While there is no guarantee the trivial irrep remains
lowest in energy for $\alpha=1$, it is our best guess for the
nearest-neighbor model.  If second neighbor exchange $J_2$ is
significant, this conclusion may change.  In particular, if $J_2$ is
larger than the effective hopping amplitude and antiferromagnetic, the
lowest energy magnon has a different symmetry.  The resulting transverse
order is more complex, with a magnetic unit at least 3 times larger than
that of the R state\cite{longpaper}. In any case, the BEC picture
implies quite generally that the symmetry for $h\gtrsim h_{c2}$ is as
low or lower than that of the R state.  Therefore, since the fully
polarized state has the full pyrochlore symmetry, with increasing field
there {\sl must} be a transition from $P4_332$ to $Fd\overline{3}m$
before the saturation field.  This may be associated with the ``kink''
observed experimentally.

The classical model of Penc {\sl et al.}\cite{Penc:prl04} successfully
explains many experimental features of \cdaf\, and \hgaf. However, 
classically, additional second and third neighbor microscopic
Heisenberg interactions are required to stabilize the R state, as
suggested by H. Ueda {\sl et al.}\cite{Ueda:prl05}.  Our results show that
quantum fluctuations generate {\sl effective} interactions that
stabilize the R state even when the microscopic exchange is purely
nearest neighbor.  Furthermore, quantum fluctuations may determine the
symmetry of the transverse spin order expected just above the plateau,
which, if measured by neutrons, would be very telling.  Experimentally,
the second and third neighbor exchanges could be determined by inelastic
neutron {\sl measurements} of the magnon spectra in the fully polarized
state, as has been done in Cs$_2$CuCl$_4$\cite{Coldea:prl02}. A more direct 
test of the relevance of quantum fluctuations would be to look for 
signatures of BEC criticality (as in
Refs.~\onlinecite{Oshikawa:99},\onlinecite{Oshikawa:jpsj04}) at the
upper edge of the plateau in \hgaf.  We note that H. Ueda {\sl et
  al.}\cite{Ueda:new} observed an {\sl increase} of $h_{c2}$ with
increasing temperature, concluding that ``thermal fluctuations stabilize
collinear spin configurations''.  The BEC picture gives an alternative
explanation: the BEC temperature grows as the magnon density
increases.

Theoretically, it is intriguing to contemplate the possibility that the
plateau phase might be close to a quantum phase transition to a 
spin liquid state.  A field-theoretic analysis, based on the mapping of
the 3:1 Hilbert space to a gauge theory with projective symmetry, indeed
predicts the R state as the simplest phase proximate to the spin
liquid\cite{longpaper}. Other calculations, predictions, and comparison
to different theoretical approaches will be made in a future
publication\cite{longpaper}.

We are very pleased to acknowledge H. Ueda, Y. Motome, M. Matsuda, H. Takagi,
and Y. Ueda. We are particularly grateful to the authors 
of Ref.~\cite{Ueda:new} for supplying us 
their experimental data prior to publications. This work 
was supported by NSF Grant DMR04-57440, PHY99-07949, 
and the Packard Foundation.

%\bibliography{paper4}

\end{document}